\documentclass{ws-ijmpcs}

\begin{document}

\markboth{P. Roig}
{Prospects for discovery of the $\tau^-\to\pi^-\ell^+\ell^-\nu_\tau$ decays}

%
\catchline{}{}{}{}{}
%

\title{Prospects for discovery of the $\tau^-\to\pi^-\ell^+\ell^-\nu_\tau$ decays}

\author{Pablo Roig}

\address{Instituto de F\'{\i}sica, Universidad Nacional Aut\'onoma de M\'exico, \\AP 20-364, M\'exico D.F. 01000, M\'exico.\\
pabloroig@fisica.unam.mx}

\maketitle


\begin{abstract}
We study the phenomenology of the $\tau^-\to\pi^-\nu_\tau\ell^+\ell^-$ decays ($\ell=e,\,\mu$), predicting the respective branching ratios and di-lepton invariant-mass spectra. 
In addition to the model-independent ($QED$) contributions, we investigate the structure-dependent ($SD$) terms, encoding features of the hadronization of $QCD$ currents. The 
relevant form factors are evaluated by supplementing Chiral Perturbation Theory with the inclusion of the lightest (axial-)vector resonance multiplet as dynamical fields. The 
Lagrangian couplings are fully predicted requiring the known $QCD$ asymptotic behavior to the relevant Green functions and associated form factors in the limit 
of an infinite number of colours. As a consequence we predict that the $\tau^-\to\pi^-\nu_\tau e^+e^-$ decays should be discovered soon while this is not granted for the 
$\ell=\mu$ case.

\keywords{Electromagnetic form factors, Hadronic tau decays; Chiral Lagrangians; Quantum Chromodynamics.}
\end{abstract}

\ccode{PACS numbers: 13.40.Gp, 13.35.Dx; 12.39.Fe; 12.38.-t.}
\vspace*{0.3cm}
The considered decays \cite{Guevara:2013wwa} have not been detected yet, although they are the crossed channels of the $\pi^-\to\ell^-e^+e^-\nu_l$ decays, measured long ago~\cite{Beringer:1900zz}. 
These decays probe the $W^\star-\gamma^\star-\pi$ vertex, with both virtual gauge bosons and complement the $\tau^-\to\pi^-\gamma\nu_\tau$ and $\pi^-\to\ell^-\gamma\nu_l$ 
decays, which are sensitive to the $W^\star-\gamma-\pi$ interaction. QCD predictions can be tested through its knowledge in the whole kinematical regime, which allows computing 
radiative corrections to the non-photon processes and the evaluation of the hadronic light-by-light contribution to the anomalous magnetic moment of the muon \cite{Toappear}. 
Generally, hadronic tau decays are very relevant to extrapolate between the known chiral limit and asymptotic regime enabling precision studies~\cite{Actis:2010gg}$^{,}$ 
\cite{Pich:2013lsa}.

The examined processes span different energy regions according to the energies of the exchanged $W$ and $\gamma$. At low momenta, Chiral Perturbation Theory \cite{ChPT} is the 
effective field theory dual to QCD and determines the low-energy (chiral) limit of the form factors. At larger energies the chiral expansion breaks down but $1/N_C$ 
provides an adequate alternative expansion parameter to enlarge the domain of applicability up to the kinematical limit. A convenient realization of these ideas for 
light-flavored mesons is Resonance Chiral Theory~\cite{RChT}, which we have followed. Its basic principles are the known chiral symmetry breaking, the discrete QCD symmetries, 
and unitary symmetry for the resonances without any ad-hoc dynamical assumption on the special role of vector mesons. Although an infinite number of states is predicted in the 
$N_C\to\infty$ limit, the $\tau^-\to\pi^-\nu_\tau\ell^+\ell^-$ decays damp completely the contribution of excited resonances canceling any dependence on the modelization of 
the spectrum. We have also included the most important leading corrections to this setting, given by tree-level local effective interactions among mesons and energy-dependent 
resonance off-shell widths~\cite{GomezDumm:2000fz}.

These processes are obtained by requiring that the photon in the one-pion radiative tau decays becomes virtual and converts into a lepton pair. As a result, analogous 
contributions are obtained: inner bremsstrahlung off the tau, off the pion or from the local $W\gamma\pi$ vertex and model dependent parts encoding the hadronization of the 
(axial-)vector current. The different interference terms are non-vanishing and sizable, in general. The hadronic form factors depend on the photon virtuality and on the 
product of photon and pion momenta. We point out that in the hadronization of the axial-vector current a diagram which vanished for on-shell photon \cite{Guo:2010dv} contributes 
in the present case, being proportional to the isovector part of the $\pi\pi$ vector form factor, for which a dispersive representation \cite{Dumm:2013zh} fulfilling analyticity 
and unitarity in the elastic region and describing data successfully was used.

Hadronic form factors must satisfy QCD short-distance behavior. By requiring it, consistent relations among the Lagrangian couplings are obtained \cite{JJYo} in agreement 
with previous results \cite{RChT}$^{,}$ \cite{Guo:2010dv}$^{,}$ \cite{Dumm:2013zh}$^{,}$ \cite{Literature} allowing to predict the phenomenology of the considered decays. A 
conservative variation of $20\%$ was allowed on these relations in order to estimate the error of the high-energy constraints \cite{Roig:2012zj}. The central values of the 
different contributions to the branching ratio of the $\tau^-\to\pi^-\nu_\tau\ell^+\ell^-$ decays ($\ell=e,\,\mu$) are displayed on the left-hand side of the table below. The 
error bands of these are given in the right-hand side of the table. The error bar of the IB contribution stems from the uncertainties on the $F_{\pi}$ decay constant and 
$\tau$ lepton lifetime \cite{Beringer:1900zz}. According to these results the $\ell=e$ decays should be discovered soon at Belle-II or at a future $\tau-c$ factory. 
On the contrary, this is not granted for the $\ell=\mu$ decays which should deserve a dedicated search at future facilities.

\begin{table*}[h!]
 \begin{center}
\begin{tabular}{|c||c|c||c|c|}
\hline
 & $\ell=e$ & $\ell=\mu$& $\ell=e$ & $\ell=\mu$\\
\hline
IB& $1.461\cdot10^{-5}$ & $1.600\cdot10^{-7}$ & $\pm 0.006\cdot10^{-5}$& $\pm 0.007\cdot10^{-7}$\\
IB-V& $-2\cdot10^{-8}$ & $1.4\cdot10^{-8}$ & $\left[-1\cdot10^{-7},1\cdot10^{-7}\right]$ & $\left[-4\cdot10^{-9},4\cdot10^{-8}\right]$\\
IB-A& $-9\cdot10^{-7}$ & $1.01\cdot10^{-7}$ & $\left[-3\cdot10^{-6},2\cdot10^{-6}\right]$ & $\left[-2\cdot10^{-7},6\cdot10^{-7}\right]$\\
VV & $1.16\cdot10^{-6}$ & $6.30\cdot10^{-7}$ & $\left[4\cdot10^{-7},4\cdot10^{-6}\right]$ & $\left[1\cdot10^{-7},3\cdot10^{-6}\right]$\\
AA& $2.20\cdot10^{-6}$ & $1.033\cdot10^{-6}$ & $\left[1\cdot10^{-6},9\cdot10^{-6}\right]$ & $\left[2\cdot10^{-7},6\cdot10^{-6}\right]$\\
V-A& $2\cdot10^{-10}$ & $-5\cdot10^{-11}$ & $\sim10^{-10}$  & $\sim10^{-10}$\\
\hline
TOTAL& $1.710\cdot10^{-5}$& $1.938\cdot10^{-6}$ & $\left(1.7^{+1.1}_{-0.3}\right)\cdot 10^{-5}$& $\left[3\cdot10^{-7},1\cdot10^{-5}\right]$\\
\hline
\end{tabular}
\label{Tab:1}
\end{center}
\end{table*}

Structure-dependent effects essentially saturate the $\ell=\mu$ decays and are $\sim15\%$ in the electron case. This implies much larger errors in the muon case, as displayed 
in the table. The normalized di-lepton invariant mass distribution in both cases is analyzed in Ref.~\cite{Guevara:2013wwa}. The inner-bremsstrahlung contribution dominates 
in either case up to $\sim0.1$ GeV$^2$ where the axial-vector contribution overtakes it. This changes the slope of the curve in a measurable way even with very limited 
statistics. With a fine binning and more events, the $\rho(770)$ contribution (through the $I=1$ pion vector form factor) will show up as a prominent peak.

These matrix elements are ready for installation in the new TAUOLA hadronic currents \cite{TAUOLA}.

The present study is also relevant for better characterizing the associated background for lepton flavour violating searches \cite{Guo:2010ny} in the 
$\tau^-\to\mu^-\ell^+\ell^-$ process.
\section*{Acknowledgments}

This work has been partially funded by Conacyt and DGAPA. The support of project PAPIIT IN106913 is also acknowledged.


\end{document}